\documentclass[twoside,a4paper]{article}
\usepackage{amsmath,graphicx,amssymb,fancyhdr,amsthm,,textcomp} 
\pdfoutput=1
\newtheorem{thm}{Theorem}[section] 
 
\newtheorem{lem}[thm]{Lemma} 
\newtheorem{prop}[thm]{Proposition} 
 
\theoremstyle{definition} 
  
\theoremstyle{remark}  
  
\def\beq{\begin{eqnarray}}  
\def\eeq{\end{eqnarray}}  
\def\bsp{\begin{split}}  
\def\esp{\end{split}}

\def\Tr{\mathrm{Tr}}  
\def\d{\mathrm{d}}

\newcommand{\mb}[1]{{\mathbb #1}}  
  
\newcommand{\mbold}[1]{\mbox{\boldmath{\ensuremath{#1}}}}

\begin{document}  
  
\title{\Large\textbf{Algebraic classification of spacetimes using discriminating 
scalar curvature invariants }}  
\author{{\large\textbf{Alan Coley$^{\heartsuit}$ and Sigbj\o rn Hervik$^{\text{\tiny\textleaf}}$} }
 \vspace{0.3cm} \\
$^{\heartsuit}$Department of Mathematics and Statistics,\\
Dalhousie University,
Halifax, Nova Scotia,\\
Canada B3H 3J5
\vspace{0.3cm}\\
$^{\text{\tiny\textleaf}}$Faculty of Science and Technology,\\  
 University of Stavanger,\\  N-4036 Stavanger, Norway   
\vspace{0.3cm} \\   
\texttt{aac@mathstat.dal.ca, ~sigbjorn.hervik@uis.no} }  
\date{\today}  
\maketitle  
\pagestyle{fancy}  
\fancyhead{} 
\fancyhead[EC]{A. Coley and S. Hervik}  
\fancyhead[EL,OR]{\thepage}  
\fancyhead[OC]{Discriminating scalar invariants}  
\fancyfoot{} 
  
\begin{abstract}   

The Weyl tensor and the Ricci tensor can be algebraically classified in a Lorentzian 
spacetime of arbitrary dimensions using alignment theory.
Used in tandem with 
the boost weight decomposition and curvature operators, the algebraic classification of
the Weyl tensor and the Ricci tensor in higher  dimensions can then
be refined utilizing their eigenbivector
and eigenvalue structure, respectively.
In particular, for a tensor of a particular algebraic type, the associated operator will have a 
restricted eigenvector structure, and this can then be used to 
determine necessary conditions for a particular algebraic type. In principle, 
this analysis can be used to study all of the
various algebraic types (and their subclasses) in more detail. 
We shall present 
an analysis of the discriminants of the associated characteristic equation
for the eigenvalues of an operator
to determine the conditions on (the associated)  curvature tensor 
for a given algebraic type.
We will describe an
algorithm which enables us to completely determine the eigenvalue structure of
the curvature operator, up to degeneracies, in terms of a set of
discriminants.
Since the characteristic equation has coefficients which can be
expressed in terms of the scalar polynomial curvature invariants
of the curvature tensor, we express these conditions 
(discriminants) in terms of these polynomial curvature invariants.
In particular, we can use the techniques decribed to study the necessary conditions in arbitrary dimensions
for the Weyl and Ricci curvature operators (and hence the higher dimensional Weyl and Ricci
tensors)  to be of algebraic type {\bf II} or {\bf D}, and create syzygies which are necessary
for the special algebraic type to be fulfilled. 
We are consequently able to determine the  necessary conditions in terms of simple
scalar polynomial curvature invariants in order for the
higher dimensional Weyl and Ricci
tensors to be of type {\bf II}  or  {\bf D}.
We explicitly determine the
scalar polynomial curvature invariants
for a Weyl or  Ricci tensor to be of type {\bf II} or {\bf D} in five dimensions.
A number of simple examples are presented to 
illustrate the calculational method and the power of the approach.
In particular, 
we will present a detailed analysis of
the important example of a 5 dimensional rotating black ring.

\end{abstract} 

\newpage

\section{Introduction}

Higher dimensional Lorentzian spacetimes are of considerable 
interest in current theoretical physics. 
Lorentzian spacetimes for which all polynomial scalar invariants 
constructed from the Riemann tensor and its covariant derivatives 
are constant are called $CSI$ spacetimes \cite{CSI4}. All curvature invariants 
of all orders vanish in an $D$-dimensional Lorentzian  $VSI$ 
spacetime \cite{Higher}. 
The higher dimensional $VSI$ and $CSI$ degenerate Kundt spacetimes are 
examples of spacetimes that are of fundamental importance since they are solutions of 
supergravity or superstring theory, when supported by appropriate 
bosonic fields \cite{CFH}. Higher dimensional black hole
solutions are also of current interest \cite{HIGHER-D-REVIEW}.

The 
introduction of {\em alignment theory} \cite{class} has made 
it possible to algebraically classify any tensor in a Lorentzian 
spacetime of arbitrary dimensions by boost weight. In particular,
the dimension-independent theory of
alignment, using the notions of an aligned null
direction and alignment order in Lorentzian geometry, 
can be applied
to the tensor
classification problem for the Weyl tensor in higher dimensions \cite{class} (thus 
generalizing the Petrov classifications 
in four dimensions (4D or $D=4$)).
 Indeed, it is
possible to categorize algebraically special tensors in terms of their
alignment type, with increasing specialization indicated by a higher
order of alignment.
In practice, a complete tensor classification in terms of alignment type is possible only 
for simple symmetry types and for small dimensions  \cite{class}.  However, partial
classification into broader categories is still desirable.
We note that alignment type suffices for the classification of 4D
Weyl tensors, but
the situation for higher-dimensional Weyl tensors is more complicated
(and different classifications in 4D are not equivalent in higher dimensions).
In the higher dimensional classification, the secondary alignment type is
also of significance.  Further refinement using bivectors is also useful (see below).

The analysis for higher
dimensional Weyl tensors can also be applied directly to the classification of
higher dimensional Riemann curvature tensors.  In particular the
higher-dimensional alignment types
give well defined categories for the Riemann tensor (although there are additional constraints coming
from the extra non-vanishing components).
We can also
use alignment to classify
the second-order symmetric Ricci tensor (which we refer to as Ricci type).
The Ricci tensor can also be classified according to 
its eigenvalue structure. In 
addition, using alignment theory, the higher dimensional Bianchi 
and Ricci identities have been computed  \cite{pravda} and a 
higher dimensional generalization of {\em Newman-Penrose 
formalism} has been presented \cite{class}. 

Another classification can be obtained by introducing {\em bivectors}
\cite{BIVECTOR}. The algebraic classification of the Weyl tensor using 
bivectors  is 
equivalent to the algebraic classification of the Weyl tensor by 
boost weight in 4D (i.e., the Petrov classification \cite{kramer}). However, these 
classifications are different in higher dimensions. In particular, 
the algebraic classification using alignment theory is rather 
course, and it may be useful to develop the algebraic 
classification of the Weyl tensor using bivectors to 
obtain a more refined classification.

The bivector formalism in 
higher  dimensional Lorentzian spacetimes was developed in \cite{BIVECTOR}.
The Weyl bivector operator was defined in a manner consistent with 
its boost weight decomposition. 
The Weyl tensor can then be algebraically classified
(based, in part, on the eigenbivector problem), which gives rise to a refinement  in dimensions
higher than four of the usual
alignment (boost weight) classification, 
in terms of the irreducible representations of the spins. 
In particular, the classification
in five dimensions was discussed in some detail \cite{BIVECTOR}.

\subsection{Scalar curvature invariants}

A {\em scalar polynomial curvature invariant of order $k$} 
(or, in short, a scalar invariant) is a scalar obtained by
contraction from a polynomial in the Riemann tensor and its
covariant derivatives up to the order $k$. The Kretschmann scalar,
$R_{abcd} R^{abcd}$,    is an example of a zeroth order invariant.
Scalar invariants have been extensively used in the study of $VSI$ and $CSI$ 
spacetimes    \cite{CSI4,Higher,CFH}.
In \cite{inv} it was proven
that a four-dimensional Lorentzian spacetime metric is either
\emph{$\mathcal{I}$-non-degenerate}, and hence completely locally
characterized by its scalar polynomial curvature invariants, 
or \emph{degenerate Kundt}.

In arbitrary dimensions, demanding that all of the zeroth polynomial Weyl invariants 
vanish implies that the Weyl type is {\bf III}, {\bf N}, or {\bf  O} (similarly 
for the Ricci type).
It would be particularly useful to find
necessary conditions in terms of simple scalar invariants in order for the
Weyl type (or the Ricci type) to be {\bf II}  or  {\bf D}. 
{\em{The main goal of this work is the determination of necessary conditions
in higher dimensions for algebraic type, and particularly type {\bf II} (or {\bf D}),  using 
scalar curvature invariants.}}

\subsubsection{Scalar invariants in 4D}

In 4D, demanding that the complex zeroth order quadratic and cubic Weyl invariants $I$ 
and $J$ vanish
($I=J=0$) implies that the Weyl (Petrov) type is {\bf III}, {\bf N}, 
or {\bf  O} \cite{Kundt}.
In addition, the Weyl tensor is
of type {\bf II} (or more special; e.g., type {\bf D}) if $27J^2= I^3$.

It is useful to express  the Weyl type {\bf II}
conditions in non-NP form. The syzygy
$I^3-27J^2=0$  is complex, whose real and
imaginary parts can be expressed using invariants of Weyl not containing
duals. The real part is equivalent to:
\begin{equation}
-11 W_{2}^3 + 33 W_2 W_4 - 18 W_6 = 0, \label{weyl1}
\end{equation}
and the imaginary part is equivalent to:
\begin{equation}
(W_{2}^2 - 2 W_4)(W_{2}^2 + W_4)^2 + 18 W_3^2(6 W_6 - 2 W_{3}^2 -
9 W_{2} W_4 + 3 W_{2}^3) = 0, \label{weyl2}
\end{equation}
where
\begin{eqnarray}
W_2 &=& \frac{1}{8}C_{abcd}C^{abcd},\\ \nonumber
W_3  &=& \frac{1}{16}C_{abcd}C^{cd}_{~~ pq}C^{pqab},\\ \nonumber
W_4 &=& \frac{1}{32}C_{abcd}C^{cd}_{~~pq}C^{pq}_{~~r s}C^{rsab}, \\ \nonumber
W_6 &=& \frac{1}{128}C_{abcd}C^{cd}_{~~pq}C^{pq}_{~~r s}C^{rs}_{~~tu}C^{tu}_{~~vw}C^{vwab}
.
\label{weyldef}
\end{eqnarray}

The Ricci type {\bf II} conditions are :

\begin{equation}
s_{3}^2(4s_{2}^3-6s_{2}s_{4}+s_{3}^2)-s_{4}^2(3s_{2}^2-4s_{4})=0,
\label{rictypeii2}
\end{equation}
where $S_{ab}$ is the trace-free Ricci tensor
$R_{ab} - \frac{1}{4} R g_{ab}$ and
{\footnote{Note the different index notation on the $s_i$ to that in \cite{Kundt},
in order to be self-consistent in this paper.}}
 
\begin{eqnarray}
s_{2} &=& \frac{1}{12}S_{a}^{~b}S_{b}^{~a},\\ \nonumber
s_{3} &=& \frac{1}{24}S_{a}^{~b}S_{b}^{~c}S_{c}^{~a},\\ \nonumber
s_{4} &=& \frac{1}{48} [S_{a}^{~b}S_{b}^{~c}S_{c}^{~d}S_{d}^{~a} -
\frac{1}{4}(S_{a}^{~b}S_{b}^{~a})^2].\label{riccidef}
\end{eqnarray}

If a
spacetime is Riemann type {\bf II}, then not only do the Weyl type {\bf II}
and Ricci type {\bf II} syzygies hold, but there are additional
alignment conditions (e.g., $C_{abcd}R^{bd},
C_{abcd}R^{be}R_{e}^{~d}$ are of type {\bf II}).

\subsection{Overview}

The Weyl tensor and the Ricci tensor can be algebraically classified in a Lorentzian 
spacetime of arbitrary dimensions by boost weight (using alignment theory).
A bivector formalism in 
higher  dimensional Lorentzian spacetimes has been developed 
to algebraically classify
the Weyl bivector operator. Used in tandem with 
the boost weight decomposition, the algebraic classification of
the Weyl tensor and the Ricci tensor (based on their eigenbivector
and eigenvalue structure, respectively) can consequently be 
refined. The purpose of this paper is the determination of necessary conditions
for the algebraic type of a higher dimensional  Weyl tensor or Ricci tensor, and 
particularly type {\bf II} (or {\bf D}),  using 
scalar curvature invariants.

For a tensor of a particular algebraic type, the associated operator \cite{OP} will have a 
restricted eigenvector structure. For a
given curvature operator in arbitrary dimensions,  we can consider the eigenvalues of this 
operator to obtain necessary conditions. In principle, the analysis can be used to study the
various subclasses in more detail. In particular, requiring the algebraic type to be {\bf II}
or {\bf D} will impose restrictions on the eigenvalues on the operator.

In this paper we shall present an analysis of the discriminants 
of the associated characteristic equation
to determine the conditions on a  tensor for a given algebraic type.
Since the characteristic equation has coefficients which can be
expressed in terms of the scalar polynomial curvature invariants
of the operator, we can consequently give conditions on the
eigenvalue structure expressed manifestly in terms of these polynomial curvature invariants.
We will describe the
algorithm which will enable us to completely determine the eigenvalue structure of
the curvature, up to degeneracies, in terms of a set of
discriminants ${}^nD_i$. The resulting syzygies (discriminants) can then be written as
special scalar  polynomial invariants.

In particular, we use the technique to study the necessary conditions in arbitrary dimensions
for the Weyl and Ricci curvature operators (and hence the higher dimensional Weyl and Ricci
tensors)  to be of algebraic type {\bf II} or {\bf D}.
We are consequently able to determine the  necessary condition(s) in terms of simple
scalar polynomial curvature invariant for the
higher dimensional Weyl and Ricci
tensors to be of type {\bf II}  or  {\bf D}.
We explicitly display the
scalar polynomial curvature invariants
for a Weyl or  Ricci tensor to be of type {\bf II} (or {\bf D}) in 5D.

A number of specific results are obtained, which are summarized at the end of the paper.
In addition, 
a number of simple examples are presented, including Einstein spaces, the
5D Schwarzschild spacetime, and
5D space with complex hyperbolic sections.
We will also present a detailed analysis of
the important example of a 5D rotating black ring \cite{RBR} which is generally of type 
${\bf I_i}$, but can also be of
type ${\bf II}$ or ${\bf D}$. This example serves to 
illustrate the calculational method and the power of the approach.
In particular, we shall show that the rotating black ring is of type 
${\bf II}$ (or type ${\bf D}$)
on the black hole horizon, by showing that a number of
discriminants (the $\mathcal{C}\mathcal{H}\mathcal{P}$ invariants) vanish there.

We briefly discuss the utility of using these methods to study
classification problems that also involve differential 
scalar polynomial curvature invariants constructed from the Riemann tensor
and its covariant derivatives, and present a simple illustration.
We also make some brief comments on possible future work. In the
Appendices we review 
the Weyl bivector operator (particularly for type {\bf II}  or  {\bf D})
and present some important
discriminants (or syzygies) that are used in the paper.

\newpage

\section{Discriminant Analysis}

We can use an analysis of the discriminants of the associated characteristic equation
to determine the conditions on a  tensor for a given algebraic type. In particular,
we shall seek necessary conditions for a higher dimensional Weyl tensor or Ricci
tensor to be type {\bf II} or {\bf D}. In principle, the analysis can be used to study the
various subclasses in more detail.

For a tensor of a particular algebraic type, the associated operator 
(acting on a vector space of dimension $n$) {\footnote{  Notation: For a Lorentzian spacetime of dimension $D$, non-capitalized 
Latin indices run 
over $1,...,D-2$, $n$ is the dimension of the vector space
(or the order of the  associated characteristic equation
for the eigenvalues) of the curvature operator  
(for the Ricci curvature tensor $n=D$ 
and for the Weyl curvature tensor
$n=D(D-1)/2$), and capitalized 
Latin letters are used to denote bivector indices.}}
will have a 
restricted eigenvector structure. For a
given curvature operator, ${\sf R}$, we can consider the eigenvalues of this operator
to obtain necessary conditions. In particular, requiring the algebraic type to be {\bf II}
or {\bf D} ({\bf II}/{\bf D}) will 
impose restrictions of the eigenvalues on the operator (e.g.,, the eigenvalue type 
(`Segre type'') will have to be of a particular form). Crucial in this discussion is the eigenvalue equation or characteristic equation \cite{OP}:
\beq
\det ({\sf R}-\lambda {\sf 1})=0. 
\eeq
This equation is a polynomial equation in $\lambda$ and the eigenvalues are the roots
of this equation.  Since the characteristic equation has coefficients which can be
expressed in terms of the invariants of ${\sf R}$, we can give conditions of the
eigenvalue structure expressed manifestly in terms of the invariants of ${\sf R}$.
Since the invariants of ${\sf R}$ are polynomial curvature invariants of spacetime,
these conditions will be referred to syzygies.  Henceforth, we will describe an
algorithm which enables us to completely determine the eigenvalue structure of
${\sf R}$ using the invariants $\Tr({\sf R}^k)$, up to degeneracies which will be
explained later.
{\footnote{ Since the
coefficients in the characteristic equations are written in terms of invariants 
of the form $\Tr({\sf R}^k)$, we do not need to consider Bianchi identities
or dimensional dependent identities to simplify the resulting polynomial expressions 
obtained.
}}

\subsection{Algorithm}
The characteristic equation can be expanded to a polynomial equation: 
\beq
f(\lambda)=\det(\lambda {\sf 1}-{\sf R})=a_0\lambda^n+a_1\lambda^{n-1}+\dots a_i\lambda^{n-i}+\dots +a_n.
\label{poly}\eeq
In our case, the coefficients are expressed in terms of invariants of ${\sf R}$, using
Newton's identities.  However, the algorithm which follows applies to any polynomial
equation.

Defining the polynomial invariants 
{\footnote{We will also use 
the notation $R1=R_1$,
since this is how it will be presented in MAPLE expressions. We have also
omitted any numerical coefficients in the definitions of the $R_i$
for convenience here (see later).}}
\beq
{\it R_1}\equiv\Tr({\sf R}),  \quad{\it R_2}\equiv\Tr({\sf R}^2), \quad {\it R_3}\equiv \Tr({\sf R}^3), \quad\text{etc}, 
\eeq
we can generally write the coefficients $a_i$ as a determinant of an $i\times i$ matrix as follows:
\beq
a_{i}=\frac{(-1)^i}{i!}\det\begin{bmatrix}
{\it R_1} & 1 & 0 & \cdots & 0 \\
{\it R_2} & {\it R_1}& 2 & \ddots & \vdots \\
{\it R_3} & {\it R_2}& {\it R_1}& \ddots & 0 \\
\vdots & \ddots & \ddots & \ddots & (i-1) \\
{\it R_i} & \dots & {\it R_3} & {\it R_2} & {\it R_1} 
\end{bmatrix},
\label{defai}\eeq
where  $a_0\equiv 1$ (and $i=1,\dots,n$). Explicitly, the first six are given by: 
\beq
a_0&=& 1,\nonumber\\
a_1&=& -{\it R_1}, \nonumber \\
a_2&=&\frac 12{\it R_1}^2-\frac 12{\it R_2}, \nonumber \\
a_3&=& -\frac 16 {\it R_1}^3 +\frac 12 {\it R_2 R_1} -\frac 13 {\it R_3}, \nonumber \\
a_4&=&\frac{1}{24}{\it R_1}^4  - \frac 14 {\it R_2 R_1}^2  + \frac 13 {\it R_3 R_1} +\frac 18 {\it R_2}^2-\frac 14 {\it R_4}, \nonumber \\
a_5&=& -\frac{1}{120}{\it R_1}^5+\frac 1{12} {\it R_2 R_1}^3-\frac 16 {\it R_3 R_1}^2-\frac 18 {\it R_1 R_2}^2 +\frac 14{\it R_4 R_1}\nonumber \\
&& +\frac 16 {\it R_2 R_3} -\frac 15 {\it R_5}.
\eeq
Also note that the order of $a_i$ is $i$; i.e., $\mathcal{O}(a_i)\sim R^i$.
{\footnote{It is often convenient to analyse the algebraic structure of the
trace-free part of the curvature operator ${\sf R}$, ${\sf S}$, where
$S_1 = 0$ and the expressions above simplify; e.g., 
$a_2=-\frac 12{\it S_2}, a_3=  -\frac 13 {\it S_3}, 
a_4= -\frac 14 {\it S_4}+\frac 18 {\it S_2}^2, \dots$.}} 

Note that if the invariant of highest order, $a_n$, is zero, i.e., $a_n=0$, then the eigenvalue equation trivially factorises and we have a zero eigenvalue. Therefore, it is convenient to first  check  the existence of zero-eigenvalues. In particular, if
\[ a_n=a_{n-1}=...=a_{n-k}=0,\] 
then there exists a zero eigenvalue of multiplicity $k+1$. If this the case then the polynomial factorises trivally and the order can be reduced. The following procedure can then be simplified accordingly. 

In general the polynomial, eq.(\ref{poly}), can be analysed and criteria for the various 'Segre types'
can be given.  The resulting syzygies are special polynomial invariants which can
be used to characterise the various eigenvalue cases; i.e., they are
\emph{discriminants}.  A complete set of discriminats can be algorithmically found and
in the following we will give the  algorithm which is found in \cite{alg} (see also \cite{Liu}).  The resulting discriminants will be
denoted ${}^nD_i$, ${}^n E_i$, ${}^n F_i$ etc., where $n$ denotes order of the
polynomial, and $i$ is a running index.  These discriminants can be given in terms of
the coefficients $a_i$; however, using Newton's identities we can express them
explicitely in terms of the polynomial invariants ${\it R_1}$, ${\it R_2}$, etc.

Given the polynomial 
\[ f(x)=a_0x^n+a_1x^{n-1}+\dots a_ix^{n-i}+\dots +a_n,\]
we define the $(2n+1)\times(2n+1)$ discrimination matrix $Disc(f)$:
\beq
\begin{bmatrix}
a_0 & a_1 & a_2 & \cdots &a_n & 0 & \cdots & 0 & 0 \\
0   &na_0 &(n-1)a_1& \cdots& a_{n-1}& 0& \cdots & 0 & 0\\ 
0   & a_0 & a_1   & \cdots & a_{n-1}&a_n&      & 0 & 0\\
0 &  0    &na_0  & \cdots & 2a_{n-2}& a_{n-1}& & 0 & 0  \\
\vdots & \vdots &   & & \vdots& \vdots & & \vdots& \vdots \\
0 & 0 &   \cdots    & 0 & na_0 & (n-1)a_1&\cdots & a_{n-1} & 0 \\
0 & 0 &   \cdots    &0  & a_0 & a_1      & \cdots& a_{n-1} & a_n
\end{bmatrix}
\eeq
Consider now the principal minor series, $\{ d_1, d_2, d_3, ...,d_{2n+1}\}$ defined as the determinants:
\beq
d_k=\det\begin{bmatrix}
\text{ the submatrix of }Disc(f) \\
\text{formed by the first} \\
\text{$k$ rows and $k$ columns}
\end{bmatrix}
\eeq
Let ${}^nD_i=d_{2i}$, $i=1,...,n$, then the \emph{discriminant sequence} of the polynomial $f(x)$ is given by 
\beq
\{{}^nD_1,{}^nD_2,{}^nD_3,...,{}^nD_n\}.
\eeq
By expressing the ${}^nD_i$ in terms of the curvature invariants, ${\it R_1}$, 
${\it R_2}$, etc, we can obtain the \emph{primary syzygies} ${}^nD_i$ for 
the operator ${\sf R}$. Note that the order of ${}^n D_i$ is 
$\mathcal{O}({}^n D_i)=R^{i(i-1)}$. 
{\footnote{To distinguish between the 
descriminants of different curvature operators (for example, the Weyl operator and the Ricci
operator) in a particular application, we shall (where necessary) add an additional
index; e.g., ${}_W^{10}D_k$, $k=1,...,10$, denote the descriminants of 
the 5D Weyl curvature operator.  }}

\paragraph{Sign List.} 
We call $[{\rm sign}({}^nD_1),{\rm sign}({}^nD_2),\dots,{\rm sign}({}^nD_n)]$, where 
\[ {\rm sign}(x)=\begin{cases} 
1, & x>0,\\
0, & x=0, \\
-1, & x<0,\\
\end{cases}\] 
the \emph{sign list} of the sequence $\{{}^nD_1,{}^nD_2,{}^nD_3,...,{}^nD_n\}$.
\paragraph{Revised Sign list.} Given a sign list $[s_1,s_2,...,s_n]$. If this contains any ``internal zeros'', i.e., if there is a subsequence $[s_i,0,0,\cdots,0,s_j]$, where $s_i\neq 0$ and $s_j\neq 0$, then we replace this subsequence with:
\[  [s_i,-s_i,-s_i,s_i,s_i,-s_i,-s_i,s_i,s_i\dots,s_j]. \]
The revised sign list will therefore contain no ``internal'' zeros, but may have zeros at the end. The revised sign list will give us the number of distinct real and complex  roots.

\paragraph{Number of real and complex roots.} 
Consider the revised sign list of $\{{}^nD_1,{}^nD_2,{}^nD_3,...,{}^nD_n\}$. Let:
\[ K=(\text{number of sign changes}), \quad L=(\text{number of non-zero members}),\] 
of the revised sign list. Then for $f(x)$:
\begin{itemize}
\item{} the number of distinct \emph{pairs of complex conjugate roots} is $K$; and
\item{} the number of distinct \emph{real roots} is $L-2K$.
\end{itemize}
If we are not interested in the multiplicities of the eigenvalues, the discriminant sequence, $\{{}^nD_1,{}^nD_2,{}^nD_3,...,{}^nD_n\}$ is sufficient. In some cases, this is enough to determine the eigenvalue structure of ${\sf R}$, but not always. 

\paragraph{Example: trace-free Ricci tensor in 3D.} 
Let ${\sf R}$ be the 3-dimensional tracefree Ricci tensor; i.e., 
{\footnote{ For 
illustrative purposes we explicitly repeat the definitions and notation here.}}
\beq
{\sf R}=(S^{\alpha}_{~\beta}), \quad S^\alpha_{~\beta}=R^{\alpha}_{~\beta}-\frac 13 R\delta^\alpha_{~\beta}.
\eeq
This implies,
\[ {\it R_1}=0, \quad {\it R_2}= S^\alpha_{~\beta}S^\beta_{~\alpha},  \quad {\it R_3}= S^\alpha_{~\beta}S^\beta_{~\delta}S^{\delta}_{~\alpha}.\] 
Consequently, 
\[ a_0=1,\quad a_1=0, \quad a_2=-\frac 12S^\alpha_{~\beta}S^\beta_{~\alpha},  
\quad a_3= -\frac 13S^\alpha_{~\beta}S^\beta_{~\delta}S^{\delta}_{~\alpha}.  \]
Using the procedure above, we get the discriminants:
\beq
{}^3D_1 &=& 3, \\
{}^3D_2 &=& -6a_2=3S^\alpha_{~\beta}S^\beta_{~\alpha}, \\
{}^3D_3 &=& -4a_2^3-27a_3^2=\frac 12(S^\alpha_{~\beta}S^\beta_{~\alpha})^3-3(S^\alpha_{~\beta}S^\beta_{~\delta}S^{\delta}_{~\alpha})^2.
\eeq
We clearly have ${}^3D_1>0$. The possible signs of the discriminants ${}^3D_2$ and ${}^3D_3$ can now be used to determine the number of real/complex eigenvalues of $S^\alpha_{~\beta}$.
\begin{enumerate}
\item{} ${}^3D_3>0$, ${}^3D_2>0$: 3 distinct real eigenvalues.
\item{} ${}^3D_3>0$, ${}^3D_2\leq 0$:  2 pairs of complex eigenvalues, which is impossible.
\item{} ${}^3D_3<0$:  1 real and 2 complex eigenvalues.
\item{}  ${}^3D_3=0$, ${}^3D_2>0$: 2 real eigenvalues (one of them must be of muliplicity 2). 
\item{} ${}^3D_3=0$, ${}^3D_2<0$: 2 complex eigenvalues (impossible, since the last eigenvalue must be real and hence, distinct).
\item{}  ${}^3D_3=0$, ${}^3D_2=0$: 1 real eigenvalue (which must be equal to zero since $S^\alpha_{~\beta}$ is tracefree). 
\end{enumerate}
Note that in terms of the Segre type, a (tracefree) Ricci type {\bf II}/{\bf D} is of type $\{21\}$, $\{(1,1)1\}$, $\{3\}$, or simpler. Consequently, if the (tracefree) Ricci is of type {\bf II}/{\bf D}, or simpler, then ${}^3D_3=0$. 

\paragraph{Multiple factor sequence.} 
We note that for polynomials of order 4 and more, the discriminants ${}^nD_i$ may 
not be sufficient to determine the complete eigenvalue structure. For example, 
for quartics, if we have 2 distinct real roots, then we cannot, using  
${}^nD_i$ only, distinguish the cases $(x-\lambda_1)^3(x-\lambda_2)$ and  
$(x-\lambda_1)^2(x-\lambda_2)^2$. Therefore, we need to go a step further 
in order to distinguish these cases. 

Consider the discriminant matrix $Disc(f)$. Define the submatrices:
\beq
M(k,l)\equiv \begin{bmatrix}
\text{ the submatrix of }Disc(f) \text{ formed by} \\
\text{the first $2k$ rows and} \\
\text{first $(2k-1)$ columns + $(2k+l)$th column}
\end{bmatrix}
\eeq
Then, construct the polynomials:
\beq
\Delta_k(f)=\sum_{i=0}^k\det[M(n-k,i)]x^{k-i},
\eeq
for $k=0,1,...,n-1$. The sequence $\{\Delta_0(f),\Delta_1(f),...,\Delta_{n-1}(f)\}$ is called the multiple factor sequence of $f(x)$ due to the following result \cite{alg}:
\begin{lem}
If the number of zeros in the revised list of the discriminant sequence 
of $f(x)$ is $k$, then $\Delta_k(f)={\rm g.c.d.}(f(x),f'(x))$. 
\end{lem}
The greatest common devisor (g.c.d.) of $f(x)$ and $f'(x)$ is thus always in the multiple factor sequence. Indeed, the polynomial $\Delta(f)\equiv {\rm g.c.d.}(f(x),f'(x))$ is the repeated part of $f(x)$, because if $\Delta (f)$ has $k$ real roots of multiplicities $n_1$, $n_2$, ...,$n_k$, and $f$ has $m$ distinct real roots, then $f$ has $k$  real roots of multiplicities $n_1+1$, $n_2+1$, ...,$n_k+1$, and $m-k$ simplie real roots (similar arguement for complex roots). Therefore, by considering $\Delta(f)$ \emph{we reduce the multiplicities of all the roots by 1}. 

We can now consider the discriminants of the polynomial $\Delta(f)$ in the same way as we computed the discriminant sequence of $f$. We will call the discriminant sequence of $\Delta(f)$ for  $\{{}^nE_1,{}^nE_2,{}^nE_3,...,{}^nE_k\}$. We can now use these to determine the sign list of the $E$-sequence, etc. We can repeat this procedure and consider $\Delta(\Delta(f))=\Delta^2(f)$, $\Delta^3(f)$ etc. These have the relation:
\beq
\Delta^{j-1}(f)&\propto &(x-\lambda_1)^{n_1+1}(x-\lambda_2)^{n_2+1}...(x-\lambda_k)^{n_k+1}(x-\lambda_{k+1})...(x-\lambda_m), \nonumber \\
\Delta^{j}(f)&\propto &(x-\lambda_1)^{n_1}(x-\lambda_2)^{n_2}...(x-\lambda_k)^{n_k},
\eeq

This gives us the following algorithm for determining the root structure (or eigenvalue structure) \cite{alg}:

\paragraph{Algorithm for Root Classification} 
\begin{enumerate}
\item{} Find the discriminant sequence of $f(x)$: 
\[  \{{}^nD_1,{}^nD_2,{}^nD_3,...,{}^nD_n\},\]
and the revised sign list. Find the number of distinct roots 
by counting sign changes and non-zero elements of the revised sign list. 
If the revised sign list contains no 0's, stop.
\item{} If the revised sign list contains $k$ zeros, then compute 
the $\Delta(f)=\Delta_k(f)$ by the definition for the multiple factor sequence. 
Then repeat step 1 for $\Delta(f)$. 
\item{} Continue considering the multiple factor sequence $\Delta^2(f)$, $\Delta^3(f)$,..., until for some $j$, the revised sign list of $\Delta^j(f)$ contains no zeros.
\item{} We now compute the number of real/complex distinct roots of $\Delta^j(f)$. 
We can now determine the roots and multiplicities of $\Delta^{j-1}(f)$, which again 
enables us to determine the roots and multiplicites of $\Delta^{j-2}(f)$ etc. 
At the end of this process, we have a complete root classification for $f(x)$.
\end{enumerate}

Note that this procedure will provide us with the discriminants (or syzygies) which
gives us a complete eigenvalue classification of \emph{any} operator ${\sf R}$.  As
explained, these discriminants can be expressed in terms of polynomials of the
invariants, ${\it R_1}=\Tr({\sf R})$, ${\it R_2}=\Tr({\sf R}^2)$, ${\it R_3}=\Tr({\sf
R}^3)$ etc., of ${\sf R}$. In principle, we can use this method to study the necessary conditions
on any curvature operator of any specific eigenvalue type.

In particular, we can use the technique to study the necessary conditions
of the Weyl and Ricci curvature operators for it to be of algebraic type {\bf II}/{\bf D}.
We note that the condition ${}^nD_n=0$ will signal a double eigenvalue since the number
of eigenvalues is maximum $(n-1)$.  If ${}^nD_{n-1}=0$ also, then we have maximum
$(n-2)$ eigenvalues, etc.  We can utilise this to create syzygies which are necessary
for the special algebraic type to be fulfilled.

\subsubsection{Type {\bf  G} and {\bf I}}

Types {\bf  G} and {\bf I} are both of equal generality with respect 
to their possible
(eigenbivector/eigenvalue) roots structure.

\subsection{Type {\bf II}/{\bf D}}
Now, for a tensor to be of type {\bf II} (or {\bf D}) then the eigenvalues of the corresponding operator need to be of a special form. Since the invariants of a type {\bf II} are the same as for type {\bf D}, we will assume type {\bf D}. 
The type {\bf D} case possesses an important symmetry, namely a boost isotropy (the tensor, not necessarily the complete spacetime). This is what we will utilise in the following.
This implies specific structure for (a) particular
discriminant(s), which then gives rise to necessary condition(s) in terms of
scalar polynomial curvature invariants.

For the Ricci tensor, we note that a type {\bf D} tensor is of Segre type $\{(1,1)11...1\}$, or simpler. This implies that the \emph{Ricci operator has at least one eigenvalue of (at least) multiplicity 2. Furthermore, all the eigenvalues are real.} 

For the Weyl tensor in $D$ dimensions we can use the bivector operator in \cite{BIVECTOR}  where the canonical form of a Weyl type {\bf D} tensor is given (see also Appendix). In particular, for type {\bf D},
\[ {\sf C}=\mathrm{blockdiag}(M,\Psi,M^t)\]
where $M$ is a $(D-2)\times(D-2)$ matrix and $\Psi$ is a square matrix (see Appendix 
for the explicit form of this matrix). Since the eigenvalues of $M$ and $M^t$ are 
the same, we have that the \emph{Weyl operator has at least $(D-2)$ eigenvalues of (at least) multiplicity 2.} 

These observations connect the algebraic types to the eigenvalue structure and enables us 
to construct the necessary discriminants.

\subsubsection{Type {\bf  II}/{\bf D} in 4D}

Applying the condition ${^3}D_3=0$ for the complex three dimensional
Weyl tensor we obtain the complex syzygy
$I^3-27J^2=0$. This is equivalent to the (12th order) real syzygies given by 
eqns. (\ref{weyl1})-(\ref{weyl2})
from the six dimensional system with ${^6}D_6=0$ and ${^6}D_5=0$.
Applying the condition ${^4}D_4=0$ for the four dimensional
trace-free Ricci  tensor we obtain the  (12th order) syzygy given by 
eqn. (\ref{rictypeii2}).
{\footnote{Note that the numerical coefficients in the definitions
of the polynomial invariants in eqns. (\ref{weyl1}) - (\ref{weyl2}) and 
(\ref{rictypeii2}) are different; for example,  ${\it R_i} \propto {\it W_i}$
(i.e., ${\it R_2} = 12 {\it W_2}$, etc., in (\ref{weyl2})).}}

We can also apply these conditions to the full Riemann tensor (to be of
type {\bf II}/{\bf D}, which implies both the Weyl and Ricci tensor are of
type {\bf II}/{\bf D} and aligned). Alternatively,
we note that these will also give us syzygies for mixed tensors.
For example requiring that Riemann tensor is type {\bf II}/{\bf D}, implies that
both Ricci and Weyl is type {\bf II}/{\bf D}, but also mixed tensors, like:  \[
L_{\mu\nu}=C_{\mu\alpha\nu\beta}R^{\alpha\beta}, \quad
M_{\mu\nu}=C_{\mu\alpha\nu\beta}R^{\alpha}_{~\delta}R^{\delta\beta}, \quad
N_{\mu\nu}=C_{\alpha\mu\lambda\pi}C^{\lambda\pi}_{~~\beta\nu}R^{\alpha\beta}.\]
The type {\bf II}/{\bf D} condition therefore implies that we have the syzygy
${}^4D_4=0$ for \emph{all of} ${\sf L}=(L^\mu_{~\nu}) $, ${\sf M}=(M^\mu_{~\nu}) $ and ${\sf
N}=(N^\mu_{~\nu})$.

\subsubsection{Type {\bf II}/{\bf D} in 5D}
For the trace-free Ricci tensor, we note that type {\bf D} has to be of Segre type $\{(1,1)111\}$ 
or simpler. This implies that 2 eigenvalues are equal, while the remaining has to 
be real. Therefore, we get the necessary (20th order) syzygy for the trace-free Ricci 
tensor to be of type {\bf II}/{\bf D}:
\[ {}^5D_5=0, \quad {}^5D_4\geq 0, \quad {}^5D_3\geq 0, \quad {}^5D_2\geq 0.\]

\paragraph{Result:} {\em{The necessary condition 
for the trace-free Ricci tensor, S, to be of algebraic type {\bf II} (or {\bf D}) in 5D 
is that the  discriminant ${}_S^5D_5$ is zero, so that the
related
scalar polynomial curvature invariant $\mathcal{D} \equiv {}_S^5D_5 = 0$}}.
{\footnote{ ~$\mathcal{D} \equiv {}_S^5D_5$ is given explicitly in  Appendix B.3.}}

For the Weyl tensor, we consider the bivector operator ${\sf C}$. 
Since the bivector space is 10-dimensional, we get a condition 
involving a syzygy of order 90! In particular, the type {\bf II} operator has 3 eigenvalues of (at least) multiplicity 2. Therefore, we get the syzygies:
\[ {}^{10}D_{10}= {}^{10}D_{9}= {}^{10}D_{8}=0.\]
Since these polynomial invariants are of particular importance, we will denote them
by
$\mathcal{C} \equiv {}_W^{10}D_{10}, \mathcal{H} \equiv {}_W^{10}D_{9}, 
\mathcal{P} \equiv {}_W^{10}D_{8}$,
the $\mathcal{C}\mathcal{H}\mathcal{P}$ Weyl invariants.

\paragraph{Result:} {\em{The necessary condition 
for the Weyl tensor to be of type {\bf II} (or {\bf D}) in 5D 
is that the 
scalar polynomial curvature invariants $\mathcal{C}=\mathcal{H}=\mathcal{P}=0$.}}

These are syzygies of order 90, 72 and 56, respectively. In principle, these can be computed, 
\footnote{ We present the expression for $\mathcal{P}$ in Appendix B.4.}
but its probably more useful to consider specific metrics\footnote{Using MAPLE
it was possible to compute some of these analytically but in practise the expressions are not very useful. However, for specific metrics these are still computable and may give useful results.}. 

A necessary condition can also be found from considering combinations 
of the Weyl tensor; for example, the operator $T^{\alpha}_{~\beta}=C^{\alpha\mu\nu\rho}C_{\beta\mu\nu\rho}$. This gives again
\[ {}_T^5{D}_5=0, \quad {}_T^5{D}_4\geq 0, \quad {}_T^5{D}_3\geq 0, \quad {}_T^5{D}_2\geq 0.\]
Note that  ${}_T^5{D}_5=0$ is now a 40th order syzygy (in the Weyl tensor;
a 20th order syzygy in the square of the Weyl tensor). {\footnote{ This syzygy is presumably
not independent of the syzygies involving  $\mathcal{C}, \mathcal{H}, \mathcal{P}$, 
which perhaps suggests that an
appropriate algebraic combination of 
$\mathcal{C}, \mathcal{H}$ and $\mathcal{P}$ either simplifies or factors. We shall return to this
in future work.}}

\subsubsection{Caveat:  A fundamental degeneracy.}  We have stressed that the
conditions determined are necessary conditions. Indeed, these conditions
may not be sufficient.  The reason for this is that the
characteristic equation for different algebraic types may be identical
and consequently the scalar invariants
are also identical.  For example, its not possible to distingush the Segre types
$\{(1,1)111\}$ (of Ricci type {\bf D}) and $\{1,11(11)\}$ (of Ricci type {\bf I}).
This implies also that the $\mathcal{C}\mathcal{H}\mathcal{P}$ conditions may be fulfilled in spite of
the fact that the spacetime is of type {\bf G} or {\bf I}.  An explicit example of this is the
following result:

\begin{prop}
Assume a 5D spacetime has a Weyl tensor with $SO(2)$ isotropy. Then it fulfills the 
$\mathcal{C}\mathcal{H}\mathcal{P}$ syzygies; i.e., $\mathcal{C}=\mathcal{H}=\mathcal{P}=0$. 
\end{prop}
This result can be seen from using the bivector operator and imposing the $SO(2)$-symmetry. One then sees that this forces that there must be 3 pairs of equal eigenvalues, which implies that there are maximum 7 distinct eigenvalues; hence, the result follows. 

This \emph{degeneracy} in the classification is a fundamental problem when
considering scalar invariants only.  Sometimes these cases can be resolved by considering
other invariants; however, there is no guarantee that this can be achieved.  Using the invariants only, we can
only determine the \emph{eigenvalue type} of the operator.  For example, if we find
the eigenvalue type to be $\{(11)111\}$, then this can correspond to three Ricci
types:  $\{(1,1)111\}$, $\{1,11(11)\}$, and $\{2111\}$ because all of these Ricci
types have the same eigenvalue type.  This is a \emph{fundamental degeneracy} in the
classification of tensors using scalar invariants only and has been discussed in earlier
papers \cite{inv}. Therefore, we need to keep this in mind when using the
discriminants.\footnote{On the other hand, the same degeneracy implies that exactly
the same procedure (and equations) can be used for spaces of other metric signatures;
that is,
the procedure to give the discriminants for other signature metrics is therefore
identical to the one given here, see \cite{OP}.}  Note that this degeneracy is a \emph{discrete} degeneracy, unlike the notion of $\mathcal{I}$-degenerate metrics \cite{inv} which require a \emph{continuous} deformation.

\subsubsection{Type {\bf II}/{\bf D} in higher dimensions} 
In higher dimensions we will obtain similar syzygies for type {\bf II}/{\bf D} 
tensors. In $n$ dimensions, the Ricci and Weyl type {\bf II}/{\bf D} conditions are the corresponding syzygies ($m=n(n-1)/2)$:
\beq
\text{Ricci:} \quad && {}^nD_n=0, \\
\text{Weyl:}\quad && {}^mD_m= {}^{m}D_{m-1}=...={}^mD_{m-n+2}=0. 
\eeq
Note that the Ricci syzygy is of order $n(n-1)$, while the highest Weyl syzygy is of order 
$n(n^2-1)(n-2)/4$.

\subsection{Examples}

\paragraph{The $\mathcal{C}\mathcal{H}\mathcal{P}$ conditions are non-trivial:}
Let us consider a simple metric which shows the $\mathcal{CHP}$ conditions are 
non-trivial conditions. Consider the solvmanifold:
\beq
\d s^2=-\d t^2+e^{2p_1 t}\d x^2+e^{2p_2 t}\d y^2+e^{2p_3 t}\d z^2+e^{2p_4 t}\d w^2
\eeq
The computation for this metric is a bit lengthy (even for MAPLE); however, 
by choosing randomly some values of the parameters\footnote{For special values 
of the parameters this metric has some symmetries; however, this is not generally 
the case.}, for example, $p_1=1$, $p_2=2$, $p_3=5$, and $p_4=-7$, MAPLE quickly calculates the values of the discriminants for the Weyl operator, obtaining: 
\beq
{}^{10}D_{10}>0,~ {}^{10}D_{9}>0,~ {}^{10}D_{8}>0,~...,{}^{10}D_{2}>0,
\eeq
showing that this metric is not type {\bf II} or {\bf D} (and that the 
$\mathcal{C}\mathcal{H}\mathcal{P}$ invariants are not trivial).

\subsubsection{Some simple examples}

\paragraph{Einstein spaces} In an Einstein space all of 
$\Tr({\sf R}^k) \propto \Lambda^k$, and hence all of the discriminants
for the trace-free Ricci eigenvalue equation are zero, and the only
non-trivial scalar invariant is the Ricci scalar $R \propto \Lambda$.

\paragraph{5D Schwarzschild spacetime:}
For the Weyl operator ${\sf C}$ we get
\[ {}^{10}D_{10}= {}^{10}D_{9}=\dots={}^{10}D_4=0, \quad {}^{10}D_3>0, \quad \quad {}^{10}D_2>0.\]
This implies that the Weyl operator has 3 distinct real eigenvalues which agrees with the results of \cite{BIVECTOR}. In fact, this spacetime is of type {\bf D}. 

\paragraph{5D space with complex hyperbolic sections.}
Let us consider the example in \cite{BIVECTOR} with complex hyperbolic spatial sections:
\beq
\d s^2= -\d t^2+a(t)^2\Big{[}e^{-2w}\left(\d x+\tfrac 12(y\d z-z\d y)\right)^2\qquad &&\nonumber \\
 +e^{-w}\left(\d y^2+\d z^2\right)+\d w^2\Big{]}.&&
\label{HC2}\eeq

For the Weyl operator ${\sf C}$ we get
\[ {}^{10}D_{10}= {}^{10}D_{9}=\dots={}^{10}D_4=0, \quad {}^{10}D_3>0, \quad \quad {}^{10}D_2>0.\]
This again implies that  the Weyl operator has 3 distinct real eigenvalues, 
like the 5D Schwarzschild spacetime. However, unlike the Schwarzschild case, we  note that the some of the invariants $a_i$, defined in eq.(\ref{defai}), are zero:
\[ a_{10}=a_9=a_8=...=a_4=0.\]
As explained, this is a signal that there is a zero-eigenvalue of multiplicity 7! Hence, since there are 3 distrinct eigenvalues of which one of must be zero with multiplicity 7, the eigenvalue structure is $\{(1111111)(11)1\}$. In particular, since the Weyl operator is trace-free, we explicitly get eigenvalues:
\[ 0~[\times 7], \quad \lambda~[\times 2], \quad -2\lambda, \]
 which agrees with the results of \cite{BIVECTOR}. 

However, this spacetime is not of type {\bf II}/{\bf D}; indeed, it is
$\mathcal{I}$-non-degenerate.  This can be seen by computing the operator
$T^{\alpha}_{~\beta}=C^{\alpha\mu\nu\rho}C_{\beta\mu\nu\rho}$ which is of ``Segre''
type $\{1,(1111)\}$.  However, due to the fundamental degeneracy in the eigenvalue
type, we need to compute differential invariants to delineate this case completely.

\subsubsection{The rotating black ring.}

The 5D rotating black ring \cite{RBR} is generally of type ${\bf I_i}$, but can also be of
type ${\bf II}$ or ${\bf D}$ at different locations and for particular values of the
parameters $\lambda, \mu$. 
Assuming that the form of the metric is given by eqn. (9) in  \cite{RBR} (in terms of the
parameters $\lambda, \mu$ , where $R$ has been set to unity), we consider the
coordinate ranges $-1 \leq x \leq 1$ and $1 \leq y < \infty$ (and hence
$0 \leq \mu \leq 1$ and  $0 \leq \lambda \leq 1$), corresponding to the
regions $B,A_2,A_3$ in  \cite{RBR} in order to retain the correct (Lorentzian) signature.
We consider the algebraic type of the 5D Weyl tensor.
Calculating the polynomial invariants $\Tr({\sf C}^k)$ and evaluating at the `target' point
$x=0$ and $y=2$ in the region under consideration, {\footnote{ Note that as 
$x  \rightarrow 1$ and $y \rightarrow 1$ we obtain flat space in this region, but
this case will not be included here.}} all of the $R_i$ and hence all of the resulting
discriminants are functions of the parameters $\lambda, \mu$ only. Then, at the
`target' point, in general the metric is of type  ${\bf I_i}$,
the case $\lambda=1$ corresponds to the Myers-Perry metric (type ${\bf D}$),
$\mu = {1}/{2}$ corresponds to the black hole horizon ($y = 1/{\mu}$,
type ${\bf II}$), $\mu = 0$ corresponds to the static subcase, and $y = 
1/{\lambda}$ corresponds to a curvature singularity.

Let us first consider the trace-free part operator $T^{\alpha}_{~\beta}=C^{\alpha\mu\nu\rho}C_{\beta\mu\nu\rho}$, which gives us the discriminant:
\beq 
{}_T^5D_5=\frac{\lambda^{12}(\lambda-\mu)^{12}(2\mu-1)^2(1-\lambda)^4(1+\lambda)^4}{(1-2\lambda)^{113}}F(\mu,\lambda)    
\eeq
where $F(\mu,\lambda)$ is a  polynomial which is generically not zero.
On the horizon $\mu=1/2$, we see that ${}_T^5D_5=0$, and computing ${}_T^5D_4$ we get ${}_T^5D_4>0$ except for special values of $\lambda$. This is a signal that the metric is of type {\bf II} on the horizon. Indeed, at the horizon, $\mu=1/2$, the computation simplifies and we can compute the $\mathcal{CHP}$ invariants:
\[ \mathcal{C}=\mathcal{H}=\mathcal{P}=0, \] 
while:
\beq
{}_W^{10}D_7&\propto&\frac{\lambda^{12}(2 \lambda^3  - \lambda^2  - 8 \lambda - 16)
    (\lambda^4  - 2\lambda^2  - 2\lambda + 1)}{(2\lambda-1)^{83}}\nonumber \\
&&\times(\lambda - 1)^2  (\lambda + 2)^2
  (\lambda + 1)^2  (4 \lambda^2  - \lambda - 6)^2  (2 \lambda^2  - \lambda - 4)^2 \nonumber \\
&& \times  (4 \lambda^3  - 4 \lambda^2  - 3 \lambda + 6)^2
 (2 \lambda^4  - 12 \lambda^3  + 9 \lambda^2  + 16 \lambda - 24)^2\nonumber \\
&&\times  (4 \lambda^4  - 14 \lambda^2  + \lambda + 12)^2 (\lambda^3  - 2 \lambda^2  + \lambda + 6)^2  (2\lambda + 1)^6,  
\eeq
where a (postive) numerical factor has been ignored.  Since, the $\mathcal{CHP}$
syzygies are satisfied, this gives further evidence that the the metric is of type {\bf
II} on the horizon.  Note that we actually get further contraints from the secondary
discriminants, as can be seen from the table in Appendix \ref{dim10}.  Indeed, by
calculating the secondary discriminant ${}_W^{10}F_3$ on the horizon, we get:
\beq
{}_W^{10}F_3&\propto & \frac{\lambda^2(\lambda^4  - 2\lambda^2  - 2\lambda + 1)(2\lambda^2-\lambda-4)^2}{(2\lambda-1)^{12}},
\eeq
which we see is non-zero as long as ${}_W^{10}D_7\neq 0$.  Consequently, as long as
${}_W^{10}D_7\neq 0$, then the eigenvalue type is $\{(11)(11)(11)1..1\}$.  This is
consitent with type {\bf II}.

Another interesting special case is $\lambda=1$ (Myers-Perry), for which both  ${}_T^5D_5= {}_T^5D_4=0$, and:
\[{}_T^5D_3=67108864\mu^2 (\mu - 1)^2  (\mu - 5)^2  (5 \mu - 1)^2  (\mu + 1)^2, \quad {}_T^5D_2>0. \]
We can also here compute the $\mathcal{CHP}$ invariants, which are all zero. 

\paragraph{Note:}
We note that  ${}_T^5{D}_5=0$ is a 40th order syzygy in the Weyl tensor. 
Therefore, a useful strategy
in practical computations (for example, determining the algebraic type
of a 5D Weyl tensor), as illustrated by this example, might be to test for necessicity using an operator
like {\sf T}, which is relatively simple. If the syzygy is not satisfied we have
a definitive result. It is  possible that the syzygy can only be satified
for certain coordinate values (or parameter values), whence the 
$\mathcal{CHP}$ syzygies can be tested in these simpler particular cases.

\newpage
\section{Conclusions}

\subsection{Discussion}
\begin{figure}[tbp]  
\centering \includegraphics[width=12cm]{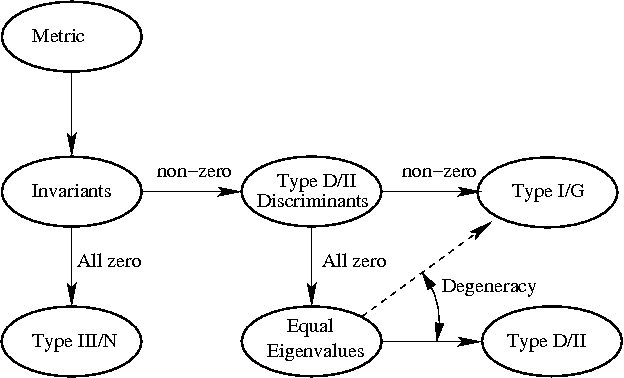}  
\caption{A flow diagram indicating how we can attempt to determine 
the algebraic type for tensors. The degeneracy indicated is due to the fact that 
several types may have the same eigenvalue type (sometimes this can be resolved by 
considering other invariants, but not always). }  
\label{fig1}  
\end{figure} 

For a curvature tensor of a particular algebraic type, the associated operator will have a 
restricted eigenvector structure. For a
given curvature operator in arbitrary dimensions,  we can thus consider the eigenvalues of this 
operator to obtain necessary conditions in order for the tensor to be of a particular 
algebraic type. 
In principle, this analysis can be used to study all of the
various subclasses of particular 
algebraic types in more detail. In particular, requiring the algebraic type 
to be {\bf II} or {\bf D}
will impose useful restrictions.
In this paper we have used an analysis of the 
discriminants of the associated characteristic equation
to determine the conditions on a  tensor for a given algebraic type.
Since the characteristic equation has coefficients which can be
expressed in terms of the scalar polynomial curvature invariants
of the operator, we can give conditions, or syzygies, on the
eigenvalue structure expressed manifestly in terms of these polynomial scalar curvature invariants.
Indeed, we have described an
algorithm which enables us to completely determine the eigenvalue structure of
the curvature, up to degeneracies, in terms of a set of
\emph{discriminants} ${}^nD_i$, ${}^nE_i$, etc.. The resulting syzygies (discriminants) can then be written as
special scalar  polynomial invariants.

In particular, we have used the technique to study the necessary conditions in arbitrary dimensions
for the Weyl and Ricci curvature operators (and hence the higher dimensional Weyl and Ricci
tensors)  to be of algebraic type {\bf II}/{\bf D}, and created syzygies which are necessary
for the special algebraic type to be fulfilled. 
We are consequently able to determine the  necessary conditions in terms of simple
scalar polynomial curvature invariant for the
higher dimensional Weyl and Ricci
tensors to be of type {\bf II}  or  {\bf D}.
We have explicitly determined the
scalar polynomial curvature invariants
for a Weyl or  Ricci tensor to be of type {\bf II} (or {\bf D}) in 5D.
This will be of considerable utility in classifying higher dimensional
solutions
obtained in
supergravity or superstring theory \cite{CFH} or higher dimensional black hole
solutions \cite{HIGHER-D-REVIEW}. 

\subsubsection{Summary of results.}

A number of specific results have been obtained in this work. 
The necessary condition 
for the trace-free Ricci tensor, $S$, to be of algebraic type {\bf II} (or {\bf D}) in 5D 
is that the  discriminant ${}_S^5D_5$ is zero, so that the
related
scalar polynomial curvature invariant $\mathcal{D} \equiv {}_S^5D_5 = 0$.
The necessary condition 
for the Weyl tensor to be of type {\bf II} (or {\bf D}) in 5D 
is that the 
scalar polynomial curvature invariants $\mathcal{C}=\mathcal{H}=\mathcal{P}=0$.
In principle, we can repeat a similar analysis for other algebraic types
(and making more use of  the secondary
discriminants).

A number of simple examples were presented, including Einstein spaces, the
5D Schwarzschild spacetime, and
5D space with complex hyperbolic sections.
In addition, a simple solvmanifold was considered to show that
the $\mathcal{C}\mathcal{H}\mathcal{P}$ conditions are non-trivial.

We also presented a detailed analysis of
the important example of a 5D rotating black ring \cite{RBR} which is generally of type 
${\bf I_i}$, but can also be of
type ${\bf II}$ or ${\bf D}$ for particular values of the
parameters. This example serves to 
illustrate the calculational method and the power of the approach.
In particular, we showed that the rotating black ring is of type 
${\bf II}$ (or type ${\bf D}$)
on the black hole horizon ($y = 1/{\mu}$), by showing that
$ \mathcal{C}=\mathcal{H}=\mathcal{P}=0$ on the horizon 
(and studying some of the secondary
discriminants). The example also illustrates the utility  
in practical computations of employing rather more simple
discriminants like ${}_T^5{D}_5=0$ (which is a 40th order 
syzygy in the Weyl tensor).

\subsection{Classification using scalar invariants}

In Lorentzian spacetimes, identical
metrics  are
often given in different coordinate systems, which
disguises their true equivalence. Perhaps the easiest way of distinguishing metrics is
through their {\em scalar polynomial curvature invariants}, due to the
fact that inequivalent invariants implies inequivalent metrics.
In \cite{inv} 
the notion of an
$\mathcal{I}$-non-degenerate spacetime metric in the class of 4D
Lorentzian manifolds, which implies that
the spacetime metric is locally determined by its scalar
polynomial curvature invariants, was introduced.  
By determining an appropriate set of projection
operators from the Riemann tensor and its covariant derivatives,
it was proven
that a 4D Lorentzian spacetime metric is either
\emph{$\mathcal{I}$-non-degenerate} or \emph{degenerate Kundt} \cite{inv}.
Therefore, a
metric that is not characterized by its curvature invariants must
be of degenerate Kundt form. 
These results were generalized to higher dimensions in
\cite{invhigher}.

\subsubsection{Differential invariants}

The $\mathcal{I}$-non-degenerate theorem
contains not only zeroth order invariants but also differential 
scalar polynomial curvature invariants constructed from the Riemann tensor
and its covariant derivatives. For example, if the spacetime is of Weyl type {\bf N}, then
the differential invariants
$\mathcal{I}_1$ and $\mathcal{I}_2$ vanish if the spacetime is degenerate Kundt \cite{pravda} (the
definitions of the invariants $\mathcal{I}_1$ and $ \mathcal{I}_2$
are given therein). Similar results follow for Weyl type {\bf III}
spacetimes (in terms of invariants $\tilde{\mathcal{I}}_1$ and
$\tilde{\mathcal{I}}_2$) and in the conformally flat (but
non-vacuum) case (in terms of similar invariants $\mathcal{I}_1$
and $\mathcal{I}_2$ constructed from the Ricci tensor \cite{pravda}.

These conditions are {\em necessary} conditions in order for a
spacetime not to be $\mathcal{I}$-non-degenerate \cite{inv}. 
In the case that $27J^2= I^3\neq 0$ (Weyl types {\bf II} or {\bf D}), in
\cite{inv} two higher order invariants $S_1$ and $S_2$ were given as
\emph{sufficient} conditions for $\mathcal{I}$-non-degeneracy (if
$27J^2= I^3$, but $S_1\neq 0$ or $S_2\neq 0$, then the spacetime
is $\mathcal{I}$-non-degenerate). {\footnote{ If the  spacetime 
is $\mathcal{I}$-non-degenerate, then essentially 
we can construct positive  boost weight terms in the derivatives
of the curvature and determine an appropriate set of differential scalar
curvature invariants.}}

This analysis in 4D can be repeated using discriminants. Let us focus on the
Ricci tensor for illustrative purposes. The necessary condition for the Ricci
tensor to be of type  {\bf II} or  {\bf D} is given by eqn. (\ref{rictypeii2})
which, as noted earlier, follows from a discrimant analysis. Now, if we consider the
covariant derivatives of the Ricci tensor, $R_{ab;cd...}$, then for the spacetime 
to be
$\mathcal{I}$-non-degenerate each covariant derivative term must also be
of type {\bf II} or  {\bf D}. Hence we could study the eigenvalue structure
of the operators associated with the $R_{ab;cd...}$ and apply the type 
{\bf II}/{\bf D} necessary conditions in turn. For example, considering the
trace-free parts of the tensors $T_{ab} = R_{ac:d}R_b^{~c;d}, R_{;ab}, \Box R_{ab}, \dots$,
we obtain necessary conditions of the form of eqn. (\ref{rictypeii2})
but with the $s_i \equiv \Tr({\sf T}^i), i=2,3,4$. {\footnote{ 
In practice it may be advantageous to work with operators involving second covariant derivatives.}}
      
This can be repeated for the Weyl tensor and in higher dimensions \cite{invhigher}.

\paragraph{Example.} The class of 
vacuum type {\bf D} spacetimes which are $\mathcal{I}$-\emph{non-degenerate},
are invariantly classified by their scalar polynomial  curvature invariants 
\cite{dim5}. 
For example, for the Kinnersley class I type {\bf D} vacuum spacetime
\cite{Kinnersley}
(the other cases work in a similar way),
there are 4 algebraically independent (complex) Cartan invariants, 
which can be written in terms of
4 independent (complex) scalar polynomial invariants
(e.g., $I$,  and invariants such as $T^{\mu}_{~\mu}$, where
\[ T^{\mu}_{~\nu} \equiv C^{\alpha\beta\gamma\delta;\mu}C_{\alpha\beta\gamma\delta;\nu}\]
which include first and second covariant 
derivatives). The Schwarzschild vacuum type {\bf D} spacetime belongs to the
Kinnersley class I and, as discussed in \cite{inv}, all of the
algebraically independent Cartan scalars
are related to the two  functionally independent polynomial
scalar 
curvature invariants $C^{\alpha\beta\gamma\delta}C_{\alpha\beta\gamma\delta}$ and 
$T^{\mu}_{~\mu}$ (which are equal to $48{M^2}{r^{-6}}$ and
$720(r-2M){M^2}{r^{-9}}$,
respectively, as functions of the two parameters $r$
and $M$ in canonical
coordinates) \cite{Ferr2}. 
The Kerr solution belongs to Kinnersley class IIA; this spacetime has been 
invariantly characterized intrinsically  \cite{Ferr1}.

\paragraph{Kerr metric.}
For illustration let us consider the example of the Kerr metric.  The Kerr metric
is of Petrov type {\bf D} and we are thus interested in whether the covariant derivative,
$\nabla C$, is of type {\bf D} or not. As noted above,
the methods discussed  earlier can also be used for differential
invariants. The covariant derivative of the Weyl tensor,
$C_{\alpha\beta\gamma\delta;\mu}$, has no natural operator associated with it.  However, we
can, for example, consider the second order operator
$T^{\mu}_{~\nu}$ defined above.
If $T^\mu_{~\nu}$ is not of type {\bf D}/{\bf II}, then $\nabla C$ cannot be of type  {\bf D}/{\bf II} either. 

For the Kerr metric, we obtain the syzygy:
\beq
{}_T^4D_4=\frac{m^{24}a^4 G^2_-G^2_+(r^2+a^2-2mr)^2(r^2+a^2\cos^2\theta-2mr)^2 \sin^4\theta}{(r^2+a^2\cos^2\theta)^{92} }f_1^2f_2, \nonumber 
\eeq
where 
\beq
G_{\pm}&=&r^4\pm 4ar^3\cos\theta-6a^2r^2\cos^2\theta\mp 4a^3r\cos^3\theta+a^4\cos^4\theta, 
\eeq
and $f_1=f_1(a,m,r,\cos\theta)$ and $f_2=f_2(a,m,r,\cos\theta)$ are some complicated 
polynomials. We note that away from the horizon, the ergosphere, and some other special 
points, this syzygy is non-zero and hence $\nabla C$ is not of type {\bf D}/{\bf II} 
(generically), outside the horizon. The Kerr metric is therefore 
$\mathcal{I}$-non-degenerate by the results of \cite{inv}.

\subsection{Physical applications}

Recently, there has been considerable interest in  
black holes in more than four dimensions \cite{HIGHER-D-REVIEW}.
While the study of black holes in higher dimensions was perhaps originally
motivated by supergravity and string theory, now
the physical properties of such black holes 
are of interest in their own right.
Indeed, studies have shown that even at the classical level gravity
in higher dimensions exhibits much richer dynamics than in 4D, and one of the
most remarkable features of higher dimensions is the non-uniqueness of 
black holes \cite{HIGHER-D-REVIEW}

There now exist a number of different higher dimensional black hole
solutions \cite{HIGHER-D-REVIEW}, including the rotating black rings \cite{RBR},
that are the subject of ongoing research in classical relativity and
string theory.
Some of these new spacetimes have be classified algebraically \cite{class,RBR}.
However, in order to make further progress it is absolutely crucial to be able
to develop new techniques
for solving the vacuum field equations in
higher dimensions and to be able to comprehensively classify such solutions,
and the algebraic techniques recently introduced \cite{class,BIVECTOR} will be of fundemental importance
in this development. However,
the algebraic techniques used to date now are rather difficult to apply, 
and the development
of simpler criteria, including the use of necessary conditions in terms
of scalar curvature invariants introduced here, will 
hopefully prove to be of great utility.

Therefore, the analysis presented in this paper will be of considerable importance for analysing
higher dimensional black hole
solutions \cite{HIGHER-D-REVIEW}
(and solutions in supergravity or superstring theory \cite{CFH}). 
Indeed, the detailed analysis of
the 5D rotating black ring \cite{RBR}
serves to 
illustrate the power of the approach.

In future work we hope to extend the analysis presented in this paper and
further generalize it to the study of differential operators. In addition, we intend to 
discuss a number of other applications, including the algebraic 
classification of some other known
higher dimensional black hole solutions.

\newpage

\section*{Acknowledgements} 

The main part of this work was done during a visit to Dalhousie
University April-June 2010 by SH. The work was supported by
NSERC of Canada (AC) and by a Leiv Eirikson mobility grant from the  
Research Council of Norway, project no: {\bf 200910/V11} (SH). 

\appendix

\section{ The Weyl Bivector operator }

Given a vector basis ${\bf k}^{\mu}$ we can define a set of (simple) bivectors 
\[ {\bf F}^A\equiv {\bf F}^{\mu\nu}={\bf F}^{[\mu\nu]}={\bf k}^{\mu}\wedge{\bf k}^{\nu}, \]
spanning the space of antisymmetric tensors of rank 2. 
Consider a $D=(2+n)$-dimensional Lorentzian space with the following null-frame $\{ {\mbold\ell}, {\bf n},{\bf m}^i\}$ so that the metric is
\[ 
\d s^2=2{\mbold\ell}{\bf n}+\delta_{ij}{\bf m}^i{\bf m}^j.
\] 
Let us consider the following bivector basis: 
\[ {\mbold\ell}\wedge{\bf m}^i, \quad  {\mbold\ell}\wedge{\bf n},\quad  {\bf m}^i\wedge{\bf m}^j,\quad   {\bf n}\wedge{\bf m}^j,\] or for short: $[0i]$, $[01]$, $[ij]$, $[1i]$. The Lorentz metric also induces a metric, $\eta_{MN}$, in bivector space. If $m=n(n-1)/2$, then
\[ (\eta_{MN})=\frac 12 \begin{bmatrix} 
0 & 0 & 0 & {\sf 1}_n \\
0 & -1& 0 & 0\\
0 & 0 & {\sf 1}_m & 0\\
{\sf 1}_n & 0 & 0 & 0
\end{bmatrix},
\]
where ${\sf 1}_n$, and ${\sf 1}_m$ are the unit matrices of size $n\times n$ and $m\times m$, respectively, and we have assumed the bivector basis is in the order given above. This metric can then be used to raise and lower bivector indices. 

Let $V\equiv \wedge^2T^*_pM$ be the vector space of bivectors at a point $p$.  Then
consider an operator ${\sf C}=(C^{~M}_{N}):  V\mapsto V$.  We will assume that it is
symmetric in the sense that $C_{MN}=C_{NM}$.
With these assumptions, the operator ${\sf C}$ can be written in the following 
$(n+1+m+n)$-block form \cite{BIVECTOR}: 
\beq\label{WeylOperator}
{\sf C}=\begin{bmatrix}
M & \hat{K} & \hat{L} & \hat{H} \\
\check{K}^t & -\Phi & -A^t & -\hat{K}^t \\
\check{L}^t & A & \bar{H} & \hat{L}^t \\
\check{H} & -\check{K} & \check{L} & M^t 
\end{bmatrix}
\eeq
Here, the block matrices $H$ (barred, checked and hatted) are all symmetric. 
Checked (hatted) matrices correspond to negative (positive) boost weight components.

The eigenbivector problem can now be formulated as follows. A bivector $F_A$ is an eigenbivector of ${\sf C}$ if and only if 
\[ C^{~M}_{N}F_M=\lambda F_N, \quad \lambda\in\mathbb{C}.\] 
Such eigenbivectors can now be determined using standard results from linear algebra. 
The Lorentz transformations (boosts, spins and null rotations)
in $(2+n)$-dimensions are explicitely written down in  \cite{BIVECTOR}.

\subsection{Weyl operator}

In particular, for the Weyl tensor we can make the following
identifications (indices $B,C,..$ should be understood as indices over $[ij]$):

\beq
\hat{H}^i_{~j}=C_{0i0j}, && \check{H}^i_{~j}=C_{1i1j},\\
\hat{L}^i_{~B}=C_{0ijk}, && \check{L}^i_{~B}=C_{1ijk}, \\
\hat{K}^i=C_{010i}, && \check{K}^i=-C_{011i}, \\
M^i_{~j}=C_{1i0j}, && \Phi=C_{0101}, \\
A^{B}=C_{01ij}, && \bar{H}^B_{~C}=C_{ijkl}.
\eeq
The Weyl tensor is also traceless and obeys the Bianchi identity: 
\[ C^{\mu}_{~\alpha\mu\beta}=0, \quad C_{\alpha(\beta\mu\nu)}=0.\] 
These conditions translate into conditions on our block matrices. 
We can consider each boost weight in turn, and use this to 
express these matrices into irreducible representations of the spins \cite{BIVECTOR}.

\subsubsection{Boost-weight 0 components}
Here we have 
\beq
C_{0101}={C_{0i1}}^i, \quad C_{0i1j}=-\tfrac 12{C_{ikj}}^k+\tfrac 12C_{01ij},\quad C_{i(jkl)}=0. 
\eeq 
 Starting with the latter, this means that the matrix $\bar{H}^B_{~C}$ fulfills the reduced Bianchi identies. It is also symmetric which means that it has the same symmetries as an $n$-dimensional Riemann tensor. Hence, we can split this into irreducible parts over $SO(n)$ using the ``Weyl tensor'', ``trace-free Ricci'' and ``Ricci scalar'' as follows ($n>2$): 
\beq
\bar{H}_{BC}&=&\bar{C}_{ijkl}+\frac{2}{n-2}\left(\delta_{i[k}\bar{R}_{l]j}-\delta_{j[k}\bar{R}_{l]i}\right)-\frac{2}{(n-1)(n-2)}\bar{R}\delta_{i[k}\delta_{l]j}, \\
\bar{R}_{ij}&=&\bar{S}_{ij}+\tfrac 1n\bar{R}\delta_{ij}.
\eeq
The remaining Bianchi identities now imply: 
\beq
M_{ij}&=&-\tfrac{1}{2n}\bar{R}\delta_{ij}-\tfrac 12 \bar{S}_{ij}-\tfrac 12A_{ij}\\
\Phi&=&-\tfrac 12\bar{R}.
\eeq
This means that the boost weight 0 components can be specified using the 
irreducible compositions above 
($\bar{R}, \bar{S}_{ij}, {A_{ij}}, 
{\bar{C}_{ijkl}}$). 
We note that in lower dimensions we have the special cases for the 
$n$-dimensional Riemann tensor: 
(i) Dim 4 ($n=2$): $\bar{S}_{ij}=\bar{C}_{ijkl}=0$,
(ii) Dim 5 ($n=3$): $\bar{C}_{ijkl}=0$,
(iii) Dim 6 ($n=4$): $\bar{C}_{ijkl}=\bar{C}^+_{ijkl}+\bar{C}^-_{ijkl}$, 
where $\bar{C}^+$ and $\bar{C}^-$ are the self-dual, and  
the anti-self-dual parts of the Weyl tensor, respectively. 
The same can be done with the antisymmetic tensor $A_{ij}=A^+_{ij}+A^-_{ij}$.

A spin $G\in SO(n)$ acts as follows on the various matrices: 
\beq
(M,\Phi,A,\bar{H})\mapsto (GMG^{-1},\Phi,\bar{G}A,\bar{G}\bar{H}\bar{G}^{-1}).
\eeq
If $C_{\mu\nu\alpha\beta}$ is the Weyl tensor, the type {\bf D} case 
is therefore completely characterised in terms 
of a $n$-dimensional Ricci tensor, a Weyl tensor, and an 
antisymmetric tensor $A_{ij}$. 
Therefore,  the spins are  first used to diagonalise the ``Ricci tensor'' 
$\bar{R}_{ij}$. This matrix can then be described in terms of the Segre-like notation 
corresponding to its eigenvalues. There is a degeneracy in the 
eigenvalues which occurs when two, or more, eigenvalues are equal. 
Using a Segre-like notation, we therefore get the types for $
\bar{R}_{ij}$:
\beq
\{1111..\},  \{(11)11..\}, \{(11)(11)...\},  \text{etc.},
\{0111..\},  \{0(11)1..\}, \{00(11)...\},  \text{etc.},
\nonumber\\
\eeq
where a zero indicates a zero-eigenvalue. 
Regarding the antisymmetric matrix $A_{ij}$, this must be of even rank
and can be put into canonical block-diagonal form, and
we can characterise an antisymmetric matrix using the rank.
The antisymmetric matrix $A$ may also possess further degeneracies.
Finally, characterisation of the ``Weyl tensor'' $\bar{C}_{ijkl}$ reduces to characterisating the Weyl tensor of the corresponding fictitious $n$-dimensional Riemannian manifold.

\subsection{The algebraic classification} 
Let us consider the classification in \cite{class}, and investigate the different
algebraic types in turn.  In general, there will be algebraically special cases 
of type {\bf G}.
The type  {\bf I}, {\bf III} and {\bf N}'s were delineated in 
\cite{BIVECTOR}. It is of interest to 
explicitly review the type {\bf II}/{\bf D}'s here.

\subsubsection{Type {\bf II}/{\bf D}}
The tensor $C_{\mu\nu\alpha\beta}$ is of type {\bf II}/{\bf D} if and only if there 
exists a null frame such that the operator ${\sf C}$ takes the form:
\beq
{\sf C}=\begin{bmatrix}
M & 0 & 0 & 0 \\
\check{K}^t & -\Phi & -A^t & 0 \\
\check{L}^t & A & \bar{H} & 0 \\
\check{H} & -\check{K} & \check{L} & M^t 
\end{bmatrix}
\eeq
For
type {\bf D} there exists a null 
frame such that, in addition, $\check{K}^t=0, \check{L}^t=0,  \check{H}=0, 
\check{K}=0, \check{L}=0$. {\footnote{ For type {\bf D} tensors, which are invariant 
under boosts, all Lorentz transformations has been utilised \emph{except} 
for the spins. }}

Then there will be algebraic subcases according to whether some of the irreducible 
components of boost weight 0 are zero or not. A complete characterisation of all such subcases 
is very involved in its full generality. 
However, a rough classification in terms of the vanishing the irreducible components 
under spins can be made: 
(a) Type {\bf II}/{\bf D}(a): $A=0$, (b)
Type {\bf II}/{\bf D}(b): $\bar{R}_{ij}=0$,
(c) Type {\bf II}/{\bf D}(c): $\bar{C}_{ijkl}=0$.
Note that we can also have a combination of these; for example, type {\bf II}(ac), 
which means that $A=0$ and $\bar{C}=0$ (i.e., further algebraically special subcases 
can arise).

\subsection{Type {\bf D} in 4D ($n=2$)}
In 4D, the Weyl operator 
can always be put into type I form by using a null rotation (hence, $\hat{H}=0$). 
Furthermore, the irreducible representations under the spins are: 
$\hat{v}^i,  \bar{R}, A, \check{v}^i, \check{H}$. 
Utilizing the
unused freedom of one spin, one boost and two null-rotations, 
in each of the algebraically special cases we can use these to simplify the Weyl 
tensor even further.

Let us only consider type {\bf D}  for illustration. For $n=2$, 
the Weyl tensor reduces to specifying two scalars, namely $\bar{R}$ and $A_{34}$. We now get
\beq
M=\begin{bmatrix}
-\tfrac 14\bar{R} & -\tfrac 12A_{34} \\
\tfrac 12A_{34} & -\tfrac 14\bar{R} 
\end{bmatrix}, \quad 
\begin{bmatrix}
-\Phi & -A^t  \\
A & \bar{H}
\end{bmatrix}=\begin{bmatrix}
\tfrac 12\bar{R} & -A_{34} \\
A_{34} & \tfrac 12\bar{R} 
\end{bmatrix};
\eeq
consequently, the Weyl operator ${\sf C}$ has eigenvalues:
\beq
\lambda_{1,2}=\lambda_{3,4}=-\frac 14(\bar{R}\pm 2iA_{34}), \quad \lambda_{5,6}=\frac 12(\bar{R}\pm 2iA_{34}).
\eeq
We note that this is in agreement with the standard type {\bf D} analysis in 4D (see \cite{kramer}). The type {\bf D} case is boost invariant, and also invariant under spins, consequently the isotropy is 2-dimensional. 

The two subcases $A_{34}=0$ and $\bar{R}=0$ (type {\bf D}(a) and {\bf D}(b), respectively) 
are in 4D referred to the purely ``electric'' and ``magnetic'' cases, 
respectively. In 4D, there is a duality relation, $\star$, which interchanges 
these two cases; i.e., $C\mapsto \star C$ interchanges the electic and magnetic parts.

\subsection{Type {\bf II}/{\bf D} in 5D ($n=3$)}
The 5D case is considerably more difficult than the 4D case. 
The complexity drastically increases and hence the number of special cases also 
increases. However, the 5D case is still managable and some 
simplifications occur (compared to the general case). Most notably, $\bar{C}_{ijkl}=0$, 
and $\check{T}^i_{jk}$ can be written, using a  matrix $\check{n}_{ij}$, as follows 
(similarly for $\hat{T}^i_{~jk}$): 
\beq 
\check{T}^i_{~jk}=\varepsilon_{jkl}\check{n}^{li},
\eeq
where the conditions on $\check{T}^i_{~jk}$ imply that $\check{n}^{ij}$ is symmetric and 
trace-free. Therefore, we can use the spins to diagonalise $\check{n}^{ij}$. Thus 
the general 
case is $\{111\}$ (all eigenvalues different), with the special 
cases $\{(11)1\}$, $\{110\}$ and $\{000\}$. Furthermore, in the general 
case, the vector $\check{v}^i$ needs not be aligned with the eigenvectors of 
$\check{n}^{ij}$. There would consequently be special cases where $\check{v}^i$ is an eigenvector of $\check{n}^{ij}$. 
The components in 5D
are displayed in Table 1.
For type {\bf D} we have that 
$\check{n}^{ij}=0$, and hence $\check{T}^i_{~jk}=0$.

\begin{table}[ht]
\begin{tabular}{|r|l|l|}
\hline 
boost weight & Ind. Components & Weyl components \\
\hline 
$+ 2$ & $\hat{H}_{ij}$  & $ C_{0i0j}=\hat{H}_{ij}$ \\
$+1$  & $\hat{v}_i$, $\hat{n}_{ij}$ & 
$C_{0ijk}=\delta_{ij}\hat{v}_k-\delta_{ik}\hat{v}_j+\varepsilon_{jkl}\hat{n}^{li}, \quad  
C_{010i}=2\hat{v}_i $ \\
$0$  & $\bar{R}$, $\bar{S}_{ij}$, $A_{ij}$ & $\begin{cases}C_{1i0j}=-\tfrac 12\bar{R}_{ij}-\tfrac 12A_{ij},\quad C_{01ij}=A_{ij},\\
C_{0101}=-\tfrac 12 \bar{R}, \quad C_{ijkl}=\bar{R}_{ijkl}
\end{cases}$ \\
$-1$ &  $\check{v}_i$, $\check{n}_{ij}$ & 
$C_{1ijk}=\delta_{ij}\check{v}_k-\delta_{ik}\check{v}_j+\varepsilon_{jkl}\check{n}^{li}, \quad  
C_{011i}=-2\check{v}_i$ \\
$-2$ &   $\check{H}_{ij}$  & $ C_{0i0j}=\check{H}_{ij}$ \\
\hline
\end{tabular}
\caption{Dimension $D=5$: 
Here $\bar{R}^k_{~ikj}=\bar{R}_{ij}=\frac 13\bar{R}\delta_{ij}+\bar{S}_{ij}$. Also see
 \cite{dim5}} 
\label{dim5}
\end{table}

\subsubsection{Type {\bf D}}
For a type {\bf D} Weyl tensor only the following components can be non-zero:
\[ \bar{R}, \quad \bar{S}^i_{~ j},\quad  A_{ij},\]
where $i,j=3,4,5$. 
Let us use the spins to diagonalise $(S^i_{~j})=\mathrm{diag}(S_{33},S_{44},S_{55})$. Without any further assumptions, the Weyl blocks take the form:
\beq
M&=&\begin{bmatrix}
-\tfrac 16\bar{R}-\tfrac 12S_{33} & -\tfrac 12A_{34} & \tfrac 12A_{53} \\
\tfrac 12A_{34} & -\tfrac 16\bar{R}- \tfrac 12S_{44} & -\tfrac 12A_{45} \\
-\tfrac 12A_{53} & \tfrac 12A_{45} & -\tfrac 16\bar{R}-\tfrac 12S_{55} 
\end{bmatrix}, \nonumber \\
\begin{bmatrix}
-\Phi & -A^t  \\
A & \bar{H}
\end{bmatrix}
&=& \begin{bmatrix}
\tfrac 12\bar{R} & -A_{45} & -A_{53} & -A_{34} \\
A_{45} & \tfrac 16\bar{R}-S_{33} & 0 & 0 \\
A_{53} & 0 &   \tfrac 16\bar{R}-S_{44} & 0 \\
A_{34} & 0 & 0 &  \tfrac 16\bar{R}-S_{55} 
\end{bmatrix}
\eeq
The general type {\bf D} tensor thus has this canonical form. 

There are two special cases where we can use the extra symmetry to get the simplified  
canonical form:
(i) $S_{33}=S_{44}=-2S_{55}$: $A_{35}=0$.
(ii) $S_{33}=S_{44}=S_{55}=0$: $A_{35}=A_{45}=0$. 
We note that case (ii) will, without further assumptions, be invariant under spatial
rotations in the $[34]$-plane (in addition to the boost).  Assuming, in addition, that
$A_{45}=0$, then case (i)  is also invariant under a rotation in the $[34]$-plane.
Assuming that $A_{ij}$ vanishes completely, we note that case (ii) enjoys the full
invariance under the spins (i.e., $SO(3)$).

\section{Some Discriminants}
For convenience, let us consider a trace-free operator ${\sf S}$ so 
that ${\it S_1}=\Tr({\sf S})=0$. We recall that ${\it S_i} \equiv \Tr({\sf S}^i)$. We also will give the table that gives the eigenvalue type using these discriminants. Here, $1_{\mb C}$ means a pair of complex conjugate eigenvalues. 

To translate into Ricci/Weyl type we need to consider degeneracies. For example, the Eigenvalue type $\{(11)11\}$, corresponds to the three Ricci types $\{(1,1)11\}$, $\{1,1(11)\}$, $\{211\}$ because all of these have the same eigenvalue type. 
\subsection{Dimension 3 Operator}
For a 3-dimensional trace-free operator (${\it S_1}=0$), the syzygies are given by:
\beq
{}^3D_2&=& 3\,{\it S_2}\nonumber \\
{}^3D_3 &=&\frac 12\,{{\it S_2}}^{3}-3\,{{\it S_3}}^{2}
\eeq
\begin{center}
\centering
\begin{tabular}{|c|c|c|c|c|c|c|}
\hline 
 ${}^3D_3$ & ${}^3D_2$  & Eigenvalue type  \\
\hline
+ & + &  $\{111\}$ \\
$-$&    &$\{1_{\mb{C}}1\}$\\ 
0 & + &    $\{(11)1\}$ \\
0 & 0 &    $\{(111)\}$ \\
\hline
\end{tabular}
\end{center}
\subsection{Dimension 4 Operator}
For a 4-dimensional trace-free operator (${\it S_1}=0$), the syzygies are given by:
\beq
{}^4D_2&=& 4\,{\it S_2}\nonumber \\
{}^4D_3 &=&-\,{{\it S_2}}^{3}+4\,{\it S_2}\,{\it S_4}-4\,{{\it S_3}}^{2}\nonumber \\
{}^4D_4&=& \frac 1{8}\,{{\it S_2}}^{6}-{\frac {5}{4}}\,{{\it S_2}}^{4}{\it S_4}-{\frac {17}{18}}\,{{\it S_3}}^{2}{{\it S_2}}^{3}\nonumber \\
&&+4\,{{\it S_2}}^{2}{{\it S_4}}^{2}+{2}\,{{\it S_3}}^{2}{\it S_2}\,{\it S_4}-\frac 13\,{{\it S_3}}^{4}-4\,{{\it S_4}}^{3}\nonumber \\
{}^4E_2&=& {{\it S_3}}^{2}+2\,{{\it S_2}}^{3}-4\,{\it S_2}\,{\it S_4}
\eeq
\begin{center}
\begin{tabular}{|c|c|c|c|c|c|c|}
\hline 
 ${}^4D_4$ & ${}^4D_3$ & ${}^4D_2$ & ${}^4E_2$ & Eigenvalue type  \\
\hline
+ & + & + &  &  $\{1111\}$ \\
+ & $\pm/0$ & $\mp/0$  & &$\{1_{\mb{C}}1_{\mb{C}}\}$   \\
$-$& $\pm/0$  & $\pm/0$  &   &$\{1_{\mb{C}}11\}$\\ 
0 & + & + &    &$\{(11)11\}$ \\
0 & $\pm$ & $\mp$  &    &$\{(11)1_{\mb{C}}\}$ \\
0 & 0 & +  & +   &  $\{(11)(11)\}$ \\
0 & 0 & +   &0  & $\{(111)1\}$ \\
0 & 0 & $-$  &    & $\{(1_{\mb{C}}1_{\mb{C}})\}$ \\
0 & 0 & 0 &  &   $\{(1111)\}$ \\
\hline
\end{tabular}
\end{center}
\subsection{Dimension 5 Operator}
For a 5-dimensional trace-free operator (${\it S_1}=0$), the syzygies are given by:
\beq
{}^5D_2&=& 5\,{\it S_2}\nonumber \\
{}^5D_3&=& -{{\it S_2}}^{3}+5\,{\it S_2}\,{\it S_4}-5\,{{\it S_3}}^{2}\nonumber\\
{}^5D_4&=& \frac 18\,{{\it S_2}}^{6}-{\frac {11}{8}}\,{{\it S_2}}^{4}{\it S_4}+{\frac {7}{24}}\,{{\it S_2}}^{3}{{\it S_3}}^{2}\nonumber\\
 && +2\,{{\it S_2}}^{2}{\it S_3}\,{\it S_5}+{\frac {19}{4}}\,{{\it S_2}}^{2}{{\it S_4}}^{2}-{\frac {61}{12}}\,{\it S_2}\,{{\it S_3}}^{2}{\it S_4}\nonumber\\
&&-5\,{\it S_2}\,{{\it S_5}}^{2}-\frac 23\,{{\it S_3}}^{4}-5\,{{\it S_4}}^{3}+10\,{\it S_3}\,{\it S_5}\,{\it S_4}\nonumber\\
{}^5D_5&=&
 \frac{21}2\,{{\it S_2}}^{2}{{\it S_3}}^{2}{{\it S_5}}^{2}-{\frac {539}{120}}\,{{\it S_2}}^{3}{{\it S_3}}^{3}{\it S_5}-{\frac {91}{72}}\,{\it S_2}\,{{\it S_3}}^{2}{{\it S_4}}^{3}-{\frac {31}{96}}\,{{\it S_2}}^{3}{{\it S_3}}^{2}{{\it S_4}}^{2}\nonumber\\
&&+{\frac {41}{96}}\,{{\it S_2}}^{5}{{\it S_3}}^{2}{\it S_4}-\frac 52\,{\it S_2}\,{{\it S_5}}^{2}{{\it S_4}}^{2}+{\frac {11}{8}}\,{{\it S_5}}^{2}{{\it S_2}}^{3}{\it S_4}-{\frac {59}{48}}\,{{\it S_3}}^{4}{\it S_4}\,{{\it S_2}}^{2}\nonumber\\
&& +{\frac {11}{48}}\,{{\it S_2}}^{6}{\it S_3}\,{\it S_5}+\frac 94\,{\it S_3}\,{\it S_5}\,{{\it S_2}}^{2}{{\it S_4}}^{2}-{\frac {31}{20}}\,{\it S_3}\,{\it S_5}\,{{\it S_2}}^{4}{\it S_4}+{\frac {4}{45}}\,{{\it S_3}}^{5}{\it S_5}\nonumber\\
&&-\frac 52\,{{\it S_3}}^{2}{\it S_4}\,{{\it S_5}}^{2}-{\frac {2}{27}}\,{{\it S_3}}^{6}{\it S_2}-{\frac {35}{3}}\,{\it S_2}\,{\it S_3}\,{{\it S_5}}^{3}+{\frac {1}{512}}\,{{\it S_2}}^{10}\nonumber\\
&& -\frac{1}{48}\,{{\it S_3}}^{4}{{\it S_4}}^{2}-{\frac {79}{400}}\,{{\it S_2}}^{5}{{\it S_5}}^{2}-{\frac {79}{1152}}\,{{\it S_2}}^{7}{{\it S_3}}^{2}\nonumber\\
&&+{\frac {151}{192}}\,{{\it S_2}}^{4}{{\it S_3}}^{4}-{\frac {7}{256}}\,{{\it S_2}}^{8}{\it S_4}+{\frac {19}{128}}\,{{\it S_2}}^{6}{{\it S_4}}^{2}+\frac 53\,{\it S_3}\,{\it S_5}\,{{\it S_4}}^{3}\nonumber\\
&&-{\frac {25}{64}}\,{{\it S_2}}^{4}{{\it S_4}}^{3}+\frac 12\,{{\it S_2}}^{2}{{\it S_4}}^{4}-\frac 14\,{{\it S_4}}^{5}+5\,{{\it S_5}}^{4}+{\frac {43}{12}}\,{{\it S_3}}^{3}{\it S_4}\,{\it S_2}\,{\it S_5}\nonumber\\
{}^5E_2&=& {\frac {91}{24}}\,{{\it S_2}}^{4}{{\it S_3}}^{2}-\frac{15}2\,{{\it S_2}}^{3}{\it S_3}\,{\it S_5}\nonumber\\
&&+{\frac {25}{4}}\,{{\it S_2}}^{2}{{\it S_5}}^{2}-{\frac {205}{12}}\,{\it S_4}\,{{\it S_2}}^{2}{{\it S_3}}^{2}+{\frac {25}{2}}\,{\it S_4}\,{\it S_2}\,{\it S_3}\,{\it S_5}-25\,{{\it S_3}}^{3}{\it S_5}\nonumber\\
&&+{\frac {25}{4}}\,{{\it S_3}}^{2}{{\it S_4}}^{2}-\frac 18\,{{\it S_2}}^{7}+{\frac {7}{8}}\,{{\it S_2}}^{5}{\it S_4}-\frac 54\,{{\it S_2}}^{3}{{\it S_4}}^{2}+{\frac {125}{6}}\,{{\it S_3}}^{4}{\it S_2}\nonumber\\
{}^5F_2&=&
 \frac 13\,{{\it S_3}}^{2}+\frac 12\,{{\it S_2}}^{3}-{\it S_2}\,{\it S_4} 
\eeq
\
\begin{center}
\begin{tabular}{|c|c|c|c|c|c|c|}
\hline 
${}^5D_5$ & ${}^5D_4$ & ${}^5D_3$ & ${}^5D_2$ & ${}^5E_2$ & ${}^5F_2$& Eigenvalue type \\
\hline
+ & + & + & + & & & $\{11111\}$ \\
+ & $\leq 0^*$ & $\leq 0^*$ &   $\leq 0^*$& & &$\{1_{\mb{C}}1_{\mb{C}}1\}$ \\
$-$&   &   &   & & &$\{1_{\mb{C}}111\}$\\ 
0 & + &   &   &  & &$\{(11)111\}$ \\
0 & $-$ &   &   &  & &$\{(11)1_{\mb{C}}1\}$ \\
0 & 0 & +  &   & 0  & & $\{(111)11\}$ \\
0 & 0 & $-$  &   &0  & &$\{(111)1_{\mb{C}}\}$ \\
0 & 0 & +  &   &$\neq 0$&  & $\{(11)(11)1\}$ \\
0 & 0 & $-$  &   &$\neq 0$&  & $\{(1_{\mb{C}}1_{\mb{C}})1\}$ \\
0 & 0 & 0 & $\neq 0$ & & $\neq 0$ &  $\{(111)(11)\}$ \\
0 & 0 & 0 & $\neq 0$ & & $ 0$ &  $\{(1111)1\}$ \\
0 & 0 & 0 & 0 & & &  $\{(11111)\}$ \\
\hline
\end{tabular}\\
$^*$ One of these conditions is sufficient.
\end{center}
\subsection{Dimension 10 Operator}
\label{dim10}
For a 10-dimensional trace-free operator (${\it S_1}=0$), we only give a partial table
indicating when the $\mathcal{C}\mathcal{H}\mathcal{P}$ invariants are zero
(we will also ignore whether the roots are real or complex). 
\begin{center}
\begin{tabular}{|c|c|c|c|c|c|c|c|}
\hline 
${}^{10}D_{10}$ & ${}^{10}D_9$ & ${}^{10}D_8$ &  ${}^{10}D_7$ & ${}^{10}E_2$ & ${}^{10}F_2$ & ${}^{10}F_3$& Eigenvalue type \\
\hline
$\neq 0$ &  &  &  & & & & $\{111..1\}$ \\
0 & $\neq 0 $ & &  &   & & &$\{(11)11...1\}$ \\
0 & 0 & $\neq 0$  &   & $\neq 0$ & & & $\{(11)(11)1...1\}$\\ 
0 & 0 & $\neq 0$  &   & 0  & & & $\{(111)1...1\}$ \\
0 & 0 & 0  & $\neq 0$  &  &   &$\neq 0$ &$\{(11)(11)(11)1..1\}$ \\
0 & 0 & 0  & $\neq 0$  &  &$\neq 0$ &  0 & $\{(111)(11)1..1\}$ \\
0 & 0 & 0  & $\neq 0$  & & 0       &   0 & $\{(1111)11..1\}$ \\

\hline
\end{tabular}\\
\end{center}

\subsubsection{The discriminant $\mathcal{P}$}

The simplest of the three syzygies, the discriminant $\mathcal{P}$ (which is of
56th order in terms of ${\it S_k}, k \leq 10$, and
contains 13377 terms) is too lengthy to write down explicitly here, {\footnote
{ The explicit expression for $\mathcal{P}$ is given in \cite{www}.}} but it has the 
symbolic form:
\beq
\mathcal{P}&=&\frac{1}{56623104000}({\it S_6S_8}-{\it S_7}^2){\it S_2}^{21}+\left( \cdots \right){\it S_2}^{20} + \nonumber \\
&& \vdots \nonumber \\
&& + \quad \cdots\quad +\quad \cdots \quad + \nonumber \\
&& \vdots \nonumber \\
&& +\left(\cdots \right){\it S_{10}}^4+(10 {\it S_4 S_2} - 10 {\it S_3}^2  - {\it S_2}^3 ) {\it S_{10}}^5.
\eeq


\newpage
\begin{thebibliography}{99}



\bibitem{CSI4} A. Coley, S. Hervik and N. 
Pelavas,  2009, Class. Quant. Grav. {\bf 26}, 125011 
[arXiv:0904.4877]; 
A. Coley, S. Hervik and N. 
Pelavas, 2006, Class. Quant. Grav. {\bf 23}, 3053;  
A. Coley, S. Hervik and N. Pelavas, 2008, Class. Quant. Grav. {\bf 25},  025008. 

 
\bibitem{Higher}    A. Coley, R. Milson, V. Pravda and A. Pravdova, 2004, 
Class. Quant. Grav. {\bf 21}, 5519 [gr-qc/0410070]. 

\bibitem{CFH} A. Coley, A. Fuster and S. Hervik, 2009, Int. J. Mod. Phys.{\bf A24}, 1119 [arXiv:0707.0957]. 
 

\bibitem{HIGHER-D-REVIEW} R. Emparan and H. S. Reall, 2008, 
 {\em Living Rev. Rel.} {\bf 11}, 6. 

\bibitem{class}  A. Coley, R. Milson, V. Pravda and A. Pravdova, 
2004, Class. Quant. Grav. {\bf 21}, L35 [gr-qc/0401008]; R. 
Milson, A. Coley, V. Pravda and A. Pravdova, 2005, Int. J. Geom. 
Meth. Mod. Phys. {\bf 2}, 41; A. Coley,  2008, Class. Quant. Grav. 
{\bf 25}, 033001 [arXiv:0710.1598]. 
 
\bibitem{pravda} V. Pravda, A. Pravdov\' a, A. Coley and R. Milson, 2004, Class. 
Quant. Grav. {\bf 21}, 2873; M. Ortaggio, V. Pravda and A. 
Pravdov\' a, 2007, Class. Quant. Grav. {\bf 24}, 1657. 
 
\bibitem{BIVECTOR}    A. Coley and S. Hervik, 2010, Class. Quant. Grav. {\bf 27}, 
015002 [arXiv:0909.1160].

 
\bibitem{kramer} H. Stephani, D. Kramer, M. A. H. MacCallum, C. A. Hoenselaers, E. Herlt, 
2003, 
\textit{Exact solutions of Einstein's field equations, second 
edition} (Cambridge University Press; Cambridge). 
 
 
\bibitem{inv}    A. Coley, S. Hervik and N. 
Pelavas, 2009, Class. Quant. Grav. {\bf 26}, 025013 [arXiv:0901.0791]. 

\bibitem{Kundt} A. Coley, S. Hervik, G. Papadopoulos and N. 
Pelavas, 2009, Class. Quant. Grav. {\bf 26}, 105016 
[arXiv:0901.0394]. 
 
\bibitem{OP}
S. Hervik and A. Coley, 2010, Class. Quant. Grav. {\bf 27}, 095014 [arXiv:1002.0505]

\bibitem{RBR} The form of the metric considered here is given in: V. Pravda and A. Pravdova, 2005, 
Gen. Rel. Grav. {\bf 37}, 1277 [gr-qc/0501003]; the
original reference is:  
R. Emparan and H. S. Reall, 2002, 
Phys. Rev. Letts. {\bf 88} 101101.

\bibitem{alg} 
L. Yang, X.R. Hou, and Z.B. Zeng, 1996, {Sci. China} Series {\bf E} {\bf 39}, 628;
L. Yang and B. Xia, {\it Explicit Criterion to Determine the Number of Positive Roots of a Polynomial}, 1997, MM Research Preprints, no.15, 134.

\bibitem{Liu}
C.-S. Liu, 2010, Comp. Phys. Comm. {\bf 181}, 317

\bibitem{invhigher} A. Coley, S. Hervik and N. 
Pelavas,  2010, Class. Quant. Grav. {\bf 27}, 102001
[arXiv1003.2373].


\bibitem{dim5} A Coley and S Hervik, 2009,
Class. Quant. Grav. {\bf{26}} 247001 [arXiv:0911.4923].

\bibitem{Kinnersley}  W. Kinnersley, 1969, J. Math. Phys. {\bf 10}, 1195.
  
\bibitem{Ferr2}  S. B. Edgar, A. Garcia-Parrado and J. M. Martin-Garcia, 2009, Class. Quant. Grav. 
{\bf 26}, 105022 [arXiv:0812.1232]; J. J. Ferrando and J. A. Saez, 1998, Class. Quant. Grav. 
{\bf 15}, 1323.

\bibitem{Ferr1} J. J. Ferrando and J. A. Saez, 2009, Class. Quant. Grav. 
{\bf 26}, 075013.

\bibitem{www} To be posted on webpage: www.ux.uis.no/$\sim$sigbjorn/PelavasInvariant.pdf


\end{thebibliography}
\end{document}